\documentclass{jfm}
\usepackage{graphicx}
\usepackage{newtxtext}
\usepackage{newtxmath}
\usepackage{natbib}
\usepackage{hyperref}
\usepackage{caption}
\hypersetup{
    colorlinks = true,
    urlcolor   = blue,
    citecolor  = red,
}

\newcommand{\RomanNumeralCaps}[1]
\linenumbers

\title{Marangoni-driven patterns, ridges, and hills in surfactant-covered parametric surface waves}

\author{Debashis~Panda\aff{1},
  Lyes~Kahouadji\aff{1},
  Laurette~S~Tuckerman\aff{2},
  Seungwon~Shin\aff{3},
  Jalel~Chergui\aff{4},
  Damir~Juric\aff{4,5}
 \and Omar~K~Matar\aff{1}\corresp{\email{o.matar@imperial.ac.uk}}}

\affiliation{\aff{1} Department of Chemical Engineering, Imperial College London, London SW7 2AZ, United Kingdom
\aff{2} Physique~et~Mécanique~des~Milieux~Hétérogènes,~CNRS,~ESPCI~Paris,~Université~PSL,~Sorbonne Université, Université de Paris, 75005 Paris, France
\aff{3} Dept. of Mechanical and System Design Engineering, Hongik Univ., Seoul 04066, Republic of Korea
\aff{4} Universit\'e Paris Saclay, Centre National de la Recherche Scientifique (CNRS), Laboratoire Interdisciplinaire des Sciences du Num\'erique (LISN), 91400 Orsay, France
\aff{5} Dept. of Applied Mathematics and Theoretical Physics, Univ. of Cambridge, Cambridge CB3 0WA,UK}

\begin{document}
\maketitle

\begin{abstract}
Parametric oscillations of an interface separating two fluid phases create nonlinear surface waves, called Faraday waves, which organise into simple patterns, like squares and hexagons, as well as complex structures, such as double hexagonal and superlattice patterns. In this work, we study the influence of surfactant-induced Marangoni stresses on the formation and transition of Faraday wave patterns. We use a quantity $B$, that assesses the relative importance of Marangoni stresses as compared to the the surface wave dynamics. Our results show that the threshold acceleration required to destabilise a surfactant-covered interface through vibration increases with increasing $B$.
For a surfactant-free interface, a square wave pattern is observed. As $B$ is incremented, we report transitions from squares to asymmetric squares, weakly wavy stripes, and ultimately to ridges and hills.  
These hills are a consequence of the bi-directional Marangoni stresses at the neck of the ridges. The mechanisms underlying the pattern transitions and the formation of exotic ridges and hills are discussed.
\end{abstract}

\section{Introduction}
\label{sec_introduction}

\cite{Faraday1831forms} noticed that vertically vibrating a fluid layer produces surface waves oscillating at half the driving frequency. Crossing a threshold amplitude, these \emph{Faraday surface waves} usually organise into patterns like squares, hexagons, triangles, and superlattices\citep{arbell2002pattern}.
Complications arise from factors such as contact line dissipation, multifrequency, and surface contamination. In this work, we focus on the effects of surface contamination on the Faraday wave patterns. 

\cite{kumar2002parametrically} presented a linear stability theory for surfactant-covered Faraday waves in the lubrication approximation. Subsequent research \citep{kumar2004faraday} emphasised the role of the phase difference that influences the Marangoni stresses. Depending on the phase difference, the Marangoni stresses may oppose (in phase) or support (out of phase) the fluid flow. \cite{ubal2005betaS, ubal2005EOS} computed the two-dimensional numerical simulations of surfactant-covered Faraday waves.  
However, these studies are limited to linearised one- or two-dimensional models, with some being carried out using lubrication theory, lacking three-dimensional studies of strongly nonlinear Marangoni effects on pattern formation in Faraday waves.  
  
\cite{perinet2009numerical} were the first to perform full three-dimensional direct numerical simulations for the study of Faraday waves. \cite{kahouadji2015numerical} further exploited the highly parallelised front tracking code, \emph{BLUE} \citep{shin2017solver}, to find supersquare patterns. 
\cite{ebo2019faraday} employed \emph{BLUE} to study  Faraday waves on a sphere. Recently, \cite{panda2023axisymmetric, panda2024drop} used the same code for studying surface waves on a water drop. \cite{shin2018jcp} further extended \emph{BLUE} by including modules to solve surfactant dynamics on the interface as well as in the bulk medium. 

In this work, we report the results of simulations of three-dimensional surfactant-covered Faraday waves; we focus on the influence of Marangoni effects on the surface wave patterns. Our study reveals that the dominance of Marangoni flow leads to transitions away from the square patterns to asymmetric squares, weakly wavy stripes, and ridges and hills. These ridges and hills are new features that occurred on a highly elastic surface. Ridges are found to rise non-uniformly and fall by forming a hill. Our direct numerical simulations help to uncover the rich physics of the dynamics of these newly observed ridges and hills.

This paper is organised as follows: First, we briefly present the problem, scaling, and the numerical method. We then present the numerical threshold acceleration which is validated by comparison with the two-dimensional simulations of \citet{ubal2005EOS}. After that, we present a phase diagram that highlights the influence of Marangoni stresses in the pattern transition of surfactant-covered Faraday waves. These patterns are analysed spectrally. Finally, we explain the newly observed ridges and hills in detail. 

\section{Problem formulation, non-dimensionalisation, and numerical method}
\begin{figure}
   \centering
    \includegraphics[width = 0.90\linewidth]{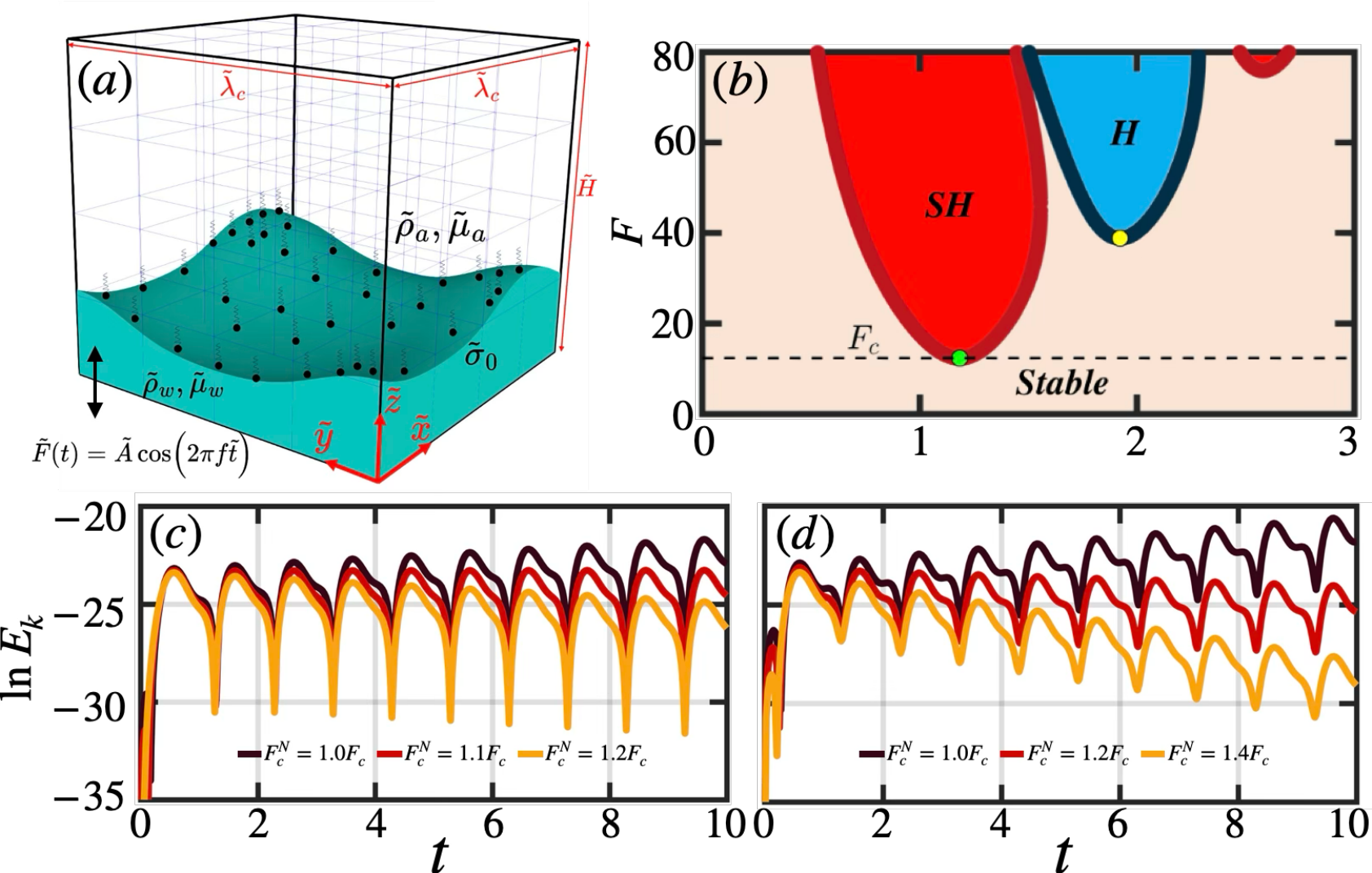}
    \caption{(a) Schematic representation of the computational domain: the height of the domain $\tilde H = 5.00~\rm mm$, and the lateral dimensions $\tilde \lambda_c \times \tilde \lambda_c$, where $\tilde \lambda_c$ is the critical wavelength. No-penetration and no-slip boundary conditions are applied at the bottom and top of the domain and periodic boundaries on the sides. (b) Critical acceleration $F_c$ for a surfactant-free interface where the solid lines represent the neutral curves for the hydrodynamic parameters listed in \cite{ubal2005EOS} and the present work, evaluated using the method of \cite{kumar1994parametric}. 'SH' and `H' refer to the subharmonic and harmonic tongues. (c,d) Temporal evolution of the total kinetic energy $E_k$ for a (c) surfactant-free and (d) surfactant-covered ($\beta_s = 1.0,~\Gamma_0 = 0.2$) interface at different acceleration amplitudes $F$. 
The wavelength in both cases is the critical wavelength $\tilde\lambda_c = 5.3023~\rm{mm}$ for the surfactant-free case
.
}\label{fig_1}
\end{figure}

Our computational domain is shown in figure \ref{fig_1}(a), which contains a layer of heavy fluid overlaid by light fluid. 
A resolution of $|\Delta x| = |\Delta y| = \lambda_c/44$ was found to be necessary to capture the Faraday wave dynamics in \cite{perinet2009numerical} and \cite{kahouadji2015numerical}. 
We choose a finer resolution of $\lambda_c/128$ to capture the coupling with the surfactant dynamics. 

We choose a simulation set-up and hydrodynamic parameters based on \cite{ubal2005EOS}, where the lower heavy fluid is a water-glycerine mixture of depth $\tilde h = 1~\rm{mm}$, density $\tilde \rho_w = 1000~\rm{kg/m^3}$, and viscosity $\tilde \mu_w = 0.025~\rm{kg/ m s }$. Unlike \cite{ubal2005EOS}, we include an upper air layer of height $4~ \rm {mm}$, density $\tilde \rho_a =1.206~\rm{kg/m^3}$, and viscosity $\tilde \mu_a = 1.82 \times 10^{-5}~\rm{kg/ m s}$.
Due to the low density ratio ($10^{-3}$) and capillary length $l_c = \sqrt{\tilde \sigma_0/\Delta\tilde \rho g } = 2.67$ mm being smaller than the air layer height, the upper fluid minimally influences the Faraday instability, allowing comparison with \cite{ubal2005EOS}. 
The surface tension of the liquid-gas surfactant-free interface is $\tilde \sigma_0 = 70 \times 10^{-3}~\rm{kg/s^2}$. The frequency of the external vibration is $100$ Hz (angular frequency $\omega=2\pi \: 100\:\rm{rad/s}$). 

We consider an insoluble surfactant that is present only on the interface since we consider that the timescale of surfactant desorption from the interface into the bulk is larger than the vibratory timescale. The saturated surfactant concentration at the critical micelle concentration is $\tilde \Gamma_\infty \sim \mathcal{O}(10^{-6})$; the range of surfactant elasticity parameter $\beta_s$ (whose definition is discussed in the following section) is $0.1 < \beta_s < 0.9$. 
The diffusivity for the surfactant $\mathcal{D}$ is set to $2.5\times 10^{-9}$ m$^2$/s to align with the work of \cite{ubal2005betaS, ubal2005EOS}. 
Unless otherwise specified, we set the initial surfactant coverage to $\tilde \Gamma_0 = 0.5\tilde \Gamma_\infty$.  

We list the major timescales in the problem: (i) the capillary timescale $\Delta \tilde t_c = \left(\tilde \rho_w \tilde h^3/\tilde \sigma_0\right)^{1/2}$ of natural capillary oscillations of the perturbed planar interface; (ii) the imposed vibrational timescale $\Delta \tilde t_i = 1/\omega$; and (iii) the Marangoni timescale $\Delta \tilde t_m = \tilde \mu_w \tilde h/(\tilde \sigma_0-\tilde \sigma(\tilde \Gamma_0))$, 
where
$\tilde\sigma$ denotes the surface tension of a surfactant-laden interface, which quantifies the surfactant dynamics on the interface. 
Our choice of parameters leads to
$\Delta \tilde t_c \sim \mathcal{O}(10^{-3})$, $\Delta \tilde t_i \sim \mathcal{O}(10^{-3})$, and $\Delta \tilde t_m \sim \mathcal{O}(10^{-4}-10^{-3})$.
This choice ensures that we observe a competition between the vibrational, capillary, and Marangoni effects. 

We choose the height of the liquid $\tilde h$ as the length scale, the inverse angular frequency $1/\omega$ as the timescale, and $\tilde \rho_w \omega^2 \tilde h^2 $ as the pressure scale. Finally, the 
interfacial concentration $\tilde \Gamma$ is scaled by the saturated interfacial concentration $\tilde \Gamma_\infty$.
The dimensionless hydrodynamic equations are then written as 
\begin{eqnarray}
 {\rho}\left(\frac{\partial {\mathbf{u}}}{\partial {t}}+ {\mathbf{u}}\cdot {\nabla}{\mathbf{u}}\right) &=& -{\nabla}{p} - \frac{\rho}{Fr^2}\left(1-F\cos{{t}}\right)\mathbf{i}_z+ \frac{1}{Re}{\nabla}\cdot \left[{\mu}({\nabla}{\mathbf{u}}+{\nabla}{\mathbf{u}}^T)\right] \nonumber\\
&& \mbox{}+ \frac{1}{We}\int_{{\mathcal{A}}({t})} \left(\sigma \kappa \mathbf{n} + {\nabla}_s {\sigma}\right) \delta({\mathbf{x}}-{\mathbf{x}}_f) d{\mathcal{A}}.
\label{eq3}
\end{eqnarray}
Here, the dimensionless density and dynamic viscosity are given by 
\begin{equation}
    \rho = \tilde \rho_a/\tilde \rho_w + (1 - \tilde \rho_a/\tilde \rho_w)\mathcal{H}({\mathbf{\tilde x}}, \tilde t), ~~~ \mu = \tilde \mu_a/\tilde \mu_w + (1 - \tilde \mu_a/\tilde \mu_w)\mathcal{H}({\mathbf{\tilde x}},\tilde t), 
\end{equation}
where $\mathcal{H}({\mathbf{\tilde x}}, \tilde t)$ is the Heaviside function, which is set to $0$ for air (subscript $a$) and $1$ for water (subscript $w$). The last term on the right-hand-side of \eqref{eq3} corresponds to the surface force at the interface ${\mathbf{x}}={\mathbf{x}}_f$. Inside the integral, 
the first and second terms account for forces arising from the normal and tangential stresses; the latter are the Marangoni stresses induced by the presence of surface tension gradients. 
$\mathcal{A}(t)$ refers to the dimensionless time-dependent interfacial area. The interfacial concentration $\tilde{\Gamma}$ evolves according to
\begin{equation}
    \frac{\partial \Gamma}{\partial t} +  \nabla_s \cdot ( \Gamma {\mathbf{u}}_s) = \frac{1}{Pe} \nabla_s^2 {\Gamma},
    \label{eq4}
\end{equation}
where $\mathbf u_s$
is the surface velocity, ${\nabla}_s$ is the gradient in the plane locally tangent to the interface. 
The dimensionless parameters in 
\eqref{eq3} and \eqref{eq4}
are the Reynolds, Weber, Peclet, and Froude numbers, and the ratio of imposed acceleration $\tilde A$ to gravitational acceration $g$:
\begin{equation}
    Re = \frac{\omega \tilde \rho_w \tilde h^2 }{\tilde \mu_w}, ~~~ We = \frac{\omega^2\tilde \rho_w \tilde h^3}{\tilde \sigma_0}, ~~~ Pe = \frac{\omega\tilde h^2}{\mathcal{D}}, ~~~ Fr = \omega \sqrt{\frac{\tilde h}{g}},~~~ F=\frac{\tilde A}{g}.
\end{equation} 
The surfactant dynamics are coupled with the hydrodynamics through the nonlinear Langmuir equation of state given by
\begin{equation}
    \sigma = \max\left[0.05,  1 + \beta_s \ln{\left(1 - {\Gamma}\right)}\right], ~~~~~
    \beta_s \equiv \frac{\mathcal{R}\tilde T\tilde \Gamma_\infty}{\tilde \sigma_0},
\label{eq:Langmuir}\end{equation}
where $\beta_s$ is the surfactant elasticity number measuring the sensitivity of the surface tension to the surfactant concentration and where the lower limit of $\sigma$ has been set to $0.05$, below which the Langmuir equation of state may diverge. 
The Marangoni stress 
$\tau$ depends on ${\Gamma}$:
\begin{equation}\label{eq:marangoni}
    \tau\equiv \frac{1}{We}{\nabla}_s {\sigma}\cdot \mathbf{t} = -\frac{Ma}{(1-\Gamma)}{\nabla}_s {\Gamma}\cdot \mathbf{t},
\end{equation}
where $Ma = \beta_s/ We$ is the Marangoni number that characterises the Marangoni strength. In the following section, however, we use a dimensionless parameter $B$ 
\begin{equation}
B \equiv 
    \frac{\tilde \sigma_0 - \tilde \sigma(\tilde \Gamma_0)}{\omega \tilde \mu_w \tilde h} = -\frac{\tilde \sigma_0 \beta_s \ln{(1-\tilde \Gamma_0/\tilde \Gamma_\infty)}}{\omega \tilde \mu_w \tilde h},
    \end{equation}
to capture the combined effect of $\beta_s$ and $\Gamma_0$ on the strength of Marangoni stresses. 

We refer 
to \cite{shin2017solver,shin2018jcp} for an exhaustive description of the numerical implementation, parallelisation and validation of the numerical framework which we briefly outline here. The spatial derivatives on the Eulerian grid are calculated using a standard cell-centered scheme, except for the nonlinear convective term for which we implemented an essentially non-oscillatory (ENO) procedure on a staggered grid. Peskin's immersed boundary method is used to couple the Eulerian and the Lagrangian grids.
The advection of the Lagrangian field $\mathbf x_f(t+\Delta t) = \int_t^{t+\Delta t} \mathbf u_f(t) dt$, where $\mathbf u_f(t)$ is the interpolated velocity at the interface at time $t$, is accomplished by second-order Runge-Kutta numerical integration. 

\section{Results and discussion}

\begin{table}
  \begin{center}
\def~{\hphantom{0}}
\caption{Numerical threshold acceleration $F_c^N(B)$ for surfactant-free and surfactant-covered interfaces 
for varying initial surfactant coverage $\Gamma_0$ and elasticity number $\beta_s$ and a fixed wavelength $\lambda_c = 5.3023$. The surfactant-free critical acceleration $F_c = 12.34$ is obtained by using the linear stability method of \cite{kumar1994parametric}. The table
demonstrates the agreement of our thresholds with those of \cite{ubal2005EOS} via $\delta^{\rm{Ubal}(B)} \equiv |F_c^N(B)-F^{\rm{Ubal}}(B)|/F_c^N(B)$.
The last column presents the increase in the Faraday threshold due to surfactant coverage via $\Delta\equiv\left(F^N_c(B)-F_c\right)/F_c$
.}
  \begin{tabular}{lccccccccc}
  \hline
  ~$\beta_s$ & ~$\Gamma_0$ & ~$B$& ~~~~Present work ($F_c^N$) & ~~$F^{\rm{Ubal}}$ & $\delta^{\rm{Ubal}}$ (\%) &~~$\Delta (\%)$\\[3 pt]
  \hline 
      clean & 0 &~0  & ~~12.32 &~ 12.30 &~0.16 &~~ 0.16\\
      1.0 & 0.1 &~0.44  & ~~13.09 &~13.00 &~0.69 & ~~ 6.07 \\
       1.0 & 0.2  &~0.89 & ~~15.45 &~15.50 &~0.32 & ~~ 25.2 \\
       1.0 & 0.3  &~1.33 & ~~18.47 &~18.51 &~0.21 & ~~ 49.7 \\
  \end{tabular}
  \captionsetup{width=1.0\textwidth}
  \label{tab_1}
  \end{center}
\end{table}
\begin{table}
  \begin{center}
\def~{\hphantom{0}}
\caption{Numerical threshold acceleration $F_c^N$ for wavelength $\lambda_c$ and varying $\beta_s$, $\Gamma_0$, and $B$, and its relative increase $\Delta\equiv(F_c^N-F_c)/F_c$ from the surfactant-free case. The highlighted data 
is used in figure \ref{fig_2}.}
  \begin{tabular}{lccccc}
  \hline
  ~~$\beta_s$~~~~ & ~$\Gamma_0$~~~~ & ~$B$~~~~  & ~$F_c^N$~~~~ &~$\Delta (\%)$~~~~\\[3 pt]
  \hline
     ~$0.10$~~~~ & ~$0.50$~~~~  & ~$0.30$~~~~ &~$13.02$~~~~ & ~$5.51$~~~~\\
     ~$0.85$~~~~ & ~$0.10$~~~~  & ~$0.40$~~~~ &~$13.09$~~~~ & ~$6.07$~~~~\\
     ~$0.25$~~~~ & ~$0.40$~~~~  & ~$0.57$~~~~ &~$14.69$~~~~ & ~$19.0$~~~~\\
     ~$0.15$~~~~ & ~$0.60$~~~~  & ~$0.61$~~~~ &~$15.32$~~~~ & ~$24.1$~~~~\\
     ~$0.20$~~~~ & ~$0.50$~~~~  & ~$0.62$~~~~ &~$15.45$~~~~ & ~$25.2$~~~~\\
     ~$0.65$~~~~ & ~$0.20$~~~~  & ~$0.65$~~~~ &~$15.62$~~~~ & ~$26.5$~~~~\\
     ~$0.75$~~~~ & ~$0.20$~~~~  & ~$0.75$~~~~ &~$15.92$~~~~ & ~$29.0$~~~~\\
     ~$0.50$~~~~ & ~$0.30$~~~~  & ~$0.79$~~~~ &~$16.05$~~~~ & ~$30.0$~~~~\\
     ~$0.35$~~~~ & ~$0.40$~~~~  & ~$0.80$~~~~ &~$16.09$~~~~ & ~$30.4$~~~~\\
\textcolor{blue}{     ~$0.30$~~~~} & \textcolor{blue}{ ~$0.50$~~~~}  & \textcolor{blue}{ ~$0.92$~~~~} &\textcolor{blue}{ ~$16.45$~~~~} & \textcolor{blue}{ ~$33.4$~~~~}\\
     ~$0.45$~~~~ & ~$0.40$~~~~  & ~$1.02$~~~~ &~$17.65$~~~~ & ~$43.0$~~~~\\
     ~$0.80$~~~~ & ~$0.25$~~~~  & ~$1.03$~~~~ &~$17.72$~~~~ & ~$43.6$~~~~\\
     ~$0.35$~~~~ & ~$0.50$~~~~  & ~$1.08$~~~~ &~$18.58$~~~~ & ~$50.5$~~~~\\
     ~$0.50$~~~~ & ~$0.40$~~~~  & ~$1.14$~~~~ &~$18.61$~~~~ & ~$50.8$~~~~\\
     \textcolor{blue}{ ~$0.40$~~~~} &\textcolor{blue}{  ~$0.50$~~~~} & \textcolor{blue}{ ~$1.23$~~~~} &\textcolor{blue}{ ~$18.63$~~~~} & \textcolor{blue}{ ~$51.0$~~~~}\\
     ~$0.60$~~~~ & ~$0.40$~~~~  & ~$1.37$~~~~ &~$18.92$~~~~ & ~$53.3$~~~~\\
     ~$0.45$~~~~ & ~$0.50$~~~~  & ~$1.39$~~~~ &~$19.01$~~~~ & ~$54.0$~~~~\\
     ~$0.35$~~~~ & ~$0.60$~~~~  & ~$1.42$~~~~ &~$19.08$~~~~ & ~$54.6$~~~~\\
     \textcolor{blue}{ ~$0.50$~~~~} & \textcolor{blue}{ ~$0.50$~~~~}  & \textcolor{blue}{ ~$1.51$~~~~} &\textcolor{blue}{ ~$19.18$~~~~} & \textcolor{blue}{ ~$55.4$~~~~}\\
     \textcolor{blue}{ ~$0.45$~~~~} & \textcolor{blue}{ ~$0.60$~~~~}  & \textcolor{blue}{ ~$1.83$~~~~} &\textcolor{blue}{ ~$20.08$~~~~} & \textcolor{blue}{ ~$62.7$~~~~}\\
     ~$0.60$~~~~ & ~$0.50$~~~~  & ~$1.85$~~~~ &~$20.83$~~~~ & ~$68.8$~~~~\\
     ~$0.70$~~~~ & ~$0.50$~~~~  & ~$2.16$~~~~ &~$20.99$~~~~ & ~$70.0$~~~~\\
       
  \end{tabular}
  \label{tab_2}
  \end{center}
\end{table}

We begin by computing the Faraday wave threshold on the surfactant-free (clean) surface using the method for linear stability analysis detailed in \citet{kumar1994parametric}. We determined that the critical acceleration amplitude $F_c$ and wavelength $\lambda_c$ are $12.34$ and $5.3023$, respectively (see figure \ref{fig_1}(b)). We can also compute a threshold from our nonlinear numerical simulations by computing the initial growth rates of the total kinetic energy $E_k$ for several values of $F$ near $F_c$. Since the growth rate varies linearly with the acceleration near the threshold, we can compute the threshold $F_c^N$ by linear interpolation. For a surfactant-free interface, we considered three acceleration amplitudes $F = (0.9,1,1.1)F_c$, as shown in figure \ref{fig_1}(b). Interpolation to zero growth rate yields $F_c^N=12.32$, which differs by only $0.16\%$ from the theoretical $F_c$, as shown in the first line of table \ref{tab_1}. 

A theoretical linear stability analysis such as that of \cite{kumar1994parametric} for a surfactant-covered interface would require linearizing the Langmuir equation of state \citep{kumar2002parametrically, kumar2004faraday}, a task that has not yet been carried out. However, we can compute the acceleration of the numerical threshold $F_c^N(B)$ using the procedure described above. We compute growth rates from numerical simulations with surfactant-covered $\beta_s = 1$ interfaces for different initial surfactant coverage $\Gamma_0$ (and corresponding values of $B$). Although the critical wavelength varies with the elasticity number \citep{kumar2004faraday}, we approximate it by its surfactant-free value. The resulting thresholds $F_c^N$ are displayed in the next three rows of table \ref{tab_1}. The same computations were carried out by  
\cite{ubal2005betaS, ubal2005EOS} using a two-dimensional finite-element technique. Their values are displayed as $F_c^{\rm Ubal}(B)$ in table \ref{tab_1}. The relative errors $\delta^{\rm Ubal}\equiv|F_c^N(B)-F_c^{\rm Ubal}(B)|/F_c^N$ between our results and those of \citet{ubal2005EOS} are less than $0.7\%$. The last column of table \ref{tab_1} shows the strong dependence of the Faraday threshold on the surfactant coverage via the relative increase $\Delta \equiv |F_c^N(B)-F_c|/F_c$. Our results show that increasing $B$ stabilises the interface, as observed in previous studies \citep{henderson1998effects, ubal2005betaS}. 

Table \ref{tab_2} shows the increase in the Faraday threshold for many other values of elasticity number $\beta_s$ and surfactant coverage $\Gamma_0$. The damping rate increases with either of these parameters, leading to an increase in the threshold of Faraday waves. We note that the threshold depends almost entirely on their combination, $B$; that is, when $\beta_s$ and $\Gamma$ are varied so as to produce the same value of $B$, then $F_c^N$ is unchanged. See, for example, the parameter pairs $(\beta_s=0.45, \Gamma_0=0.40)$, which yield $B=1.02$ $F_c^N=43.0$ and $(\beta_s=0.80,\Gamma_0=0.25)$, which yield $B=1.03$ and $F_c^N=43.6$. Other pairs of $(\beta,\Gamma_0)$ values that yield very close values of $B$ and $F_c^N$ can also be seen in table \ref{tab_2}.

After a transient phase, Faraday waves appear, which correspond to subharmonic waves whose amplitude is steady and whose response period $T$ is twice that of the forcing period. We set $t=0$ to be an instant within the steady-amplitude Faraday-wave regime. The computations for assessing the influence of $B$ on the interfacial dynamics in the nonlinear regime are then carried out for $F = 1.1F_c^N$ for which square patterns are observed in the surfactant-free case.

\begin{figure}
   \centering
    \includegraphics[width = 0.95\linewidth]{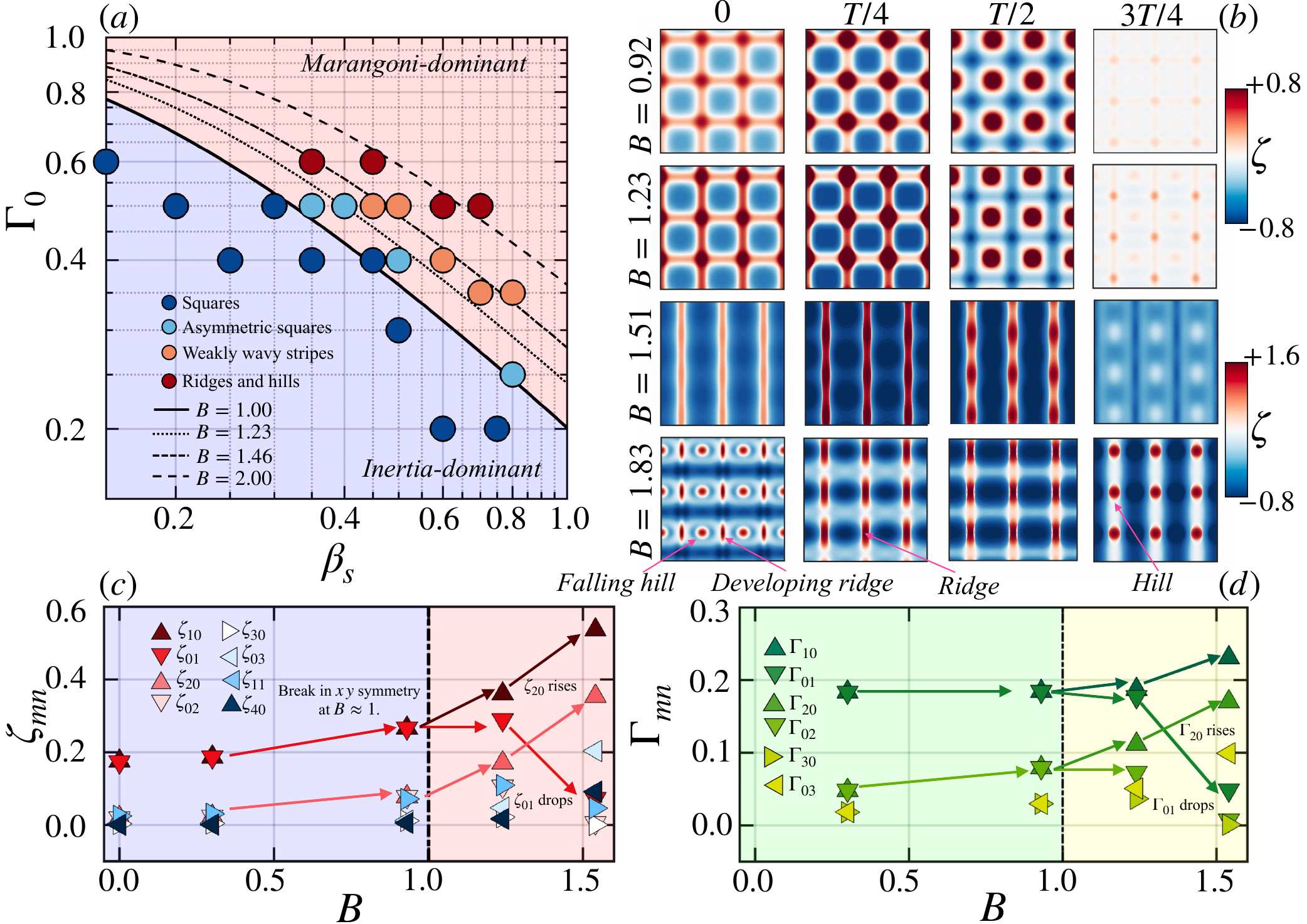}
    \caption{(a)
    Phase diagram in the $\beta_s-\Gamma_0$ parameter plane  showing the inertia-dominated (violet) and Marangoni-dominated (pink) regions. The solid, dotted, dot-dashed, and dashed lines correspond to the $B = 1, 1.23, 1.46, 2$ contours, respectively. The four typical patterns are squares, asymmetric squares, weakly wavy stripes, and ridges and hills. The phase boundaries are accurate to within $\Delta B = \pm 0.1$. The corresponding values of $B$ and $F_c^N$ are reported in table II. (b) Spatiotemporal evolution of the surface deflection $\zeta$ over one time period is shown  from left to right; squares ($B = 0.92$), asymmetric squares ($B = 1.23$), weakly wavy stripes ($B = 1.51$), and ridges and hills ($B = 1.83$) are shown from top to bottom rows, respectively. (c,d) $\zeta_{mn}$ and $\Gamma_{mn}$, the maximal magnitudes over time of the $\zeta$ and $\Gamma$ Fourier coefficients, respectively, as a function of $B$.
}\label{fig_2}
\end{figure}

As shown in figure\ \ref{fig_2}(a), for $B < 1$ (dark blue dots, purple region), the interface exhibits square symmetry. In a narrow band of $1\le B\le 1.23$ (light blue dots), the vertical and horizontal directions differ slightly; we refer to these patterns as \textit{asymmetric squares}. Within $1.23\le B \le1.46$ (orange dots), the asymmetric square pattern undergoes a transition to \textit{weakly wavy stripes}. Ridges (ellipses whose major axes are in the $y$-direction) appear very faintly as dots for $B = 1.23$, $t = 3T/4$, and more prominently on the wavy stripes for $B=1.51$. For $B=1.83$, $t=0$, one can also see circular hills between each set of ridges. The hills are the continuation of the ridges formed in the previous half-period. One such instance is shown at $t = 3T/4$, where the ridges have disappeared but the hills are present. We explore below the role of Marangoni stresses in the formation of these patterns. 
\begin{figure}    
\includegraphics[width = 0.95\linewidth]{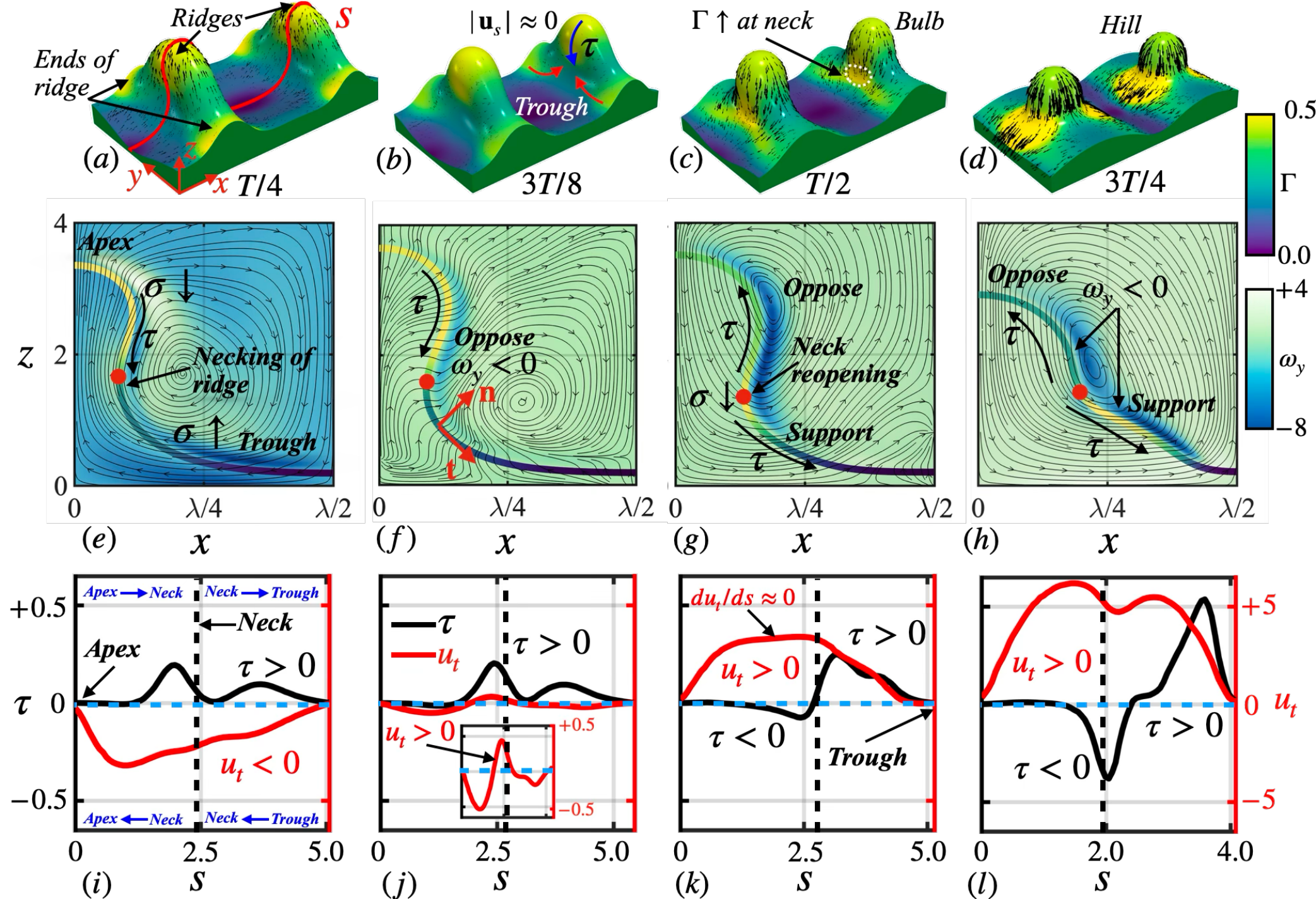}
    \centering
    \caption{(a)-(d) Three-dimensional visualization of the surface. (a) Rise of ridges and necking process at $t = T/4$ and (b)  maximum rise of the ridge at $t = 3T/8$. (c) Prominent hill on the ridge at $t = T/2$. (d) Falling hill
    at $t = 3T/4$. (e-h) Two-dimensional projections on $x-z$ slice containing interface curve $s$ (indicated in (a)) for $t = T/4, 3T/8, T/2,$ and $3T/4$, respectively. A half-wavelength (ridge to trough) is shown. Color-coding of the plane indicates $y-$vorticity $\omega_y$, while streamlines show flow in $x-z$ plane. The interface curve $s$ is colored according to the surfactant concentration. Red dots indicate the point of maximum curvature. (i-l) Tangential (see (f)) Marangoni stress and velocity along $s$ at $t = T/4, 3T/8, T/2, $ and $3T/4$ shown as black and red curves, respectively.  
    When the sign of one of these quantities is positive (negative), its direction points rightwards (leftwards) from the apex (trough) through the neck to the trough (apex) of the ridge, as indicated at the top (bottom) of figure \ref{fig2}(i).
    The vertical dashed line indicates the necking region, shown as the red dot in the corresponding $x-z$ projection. The length of $s$ decreases from about $5$ at $t=T/4, 3T/8,$ $T/2$ to about $4$ at $t=3T/4$, as can be seen in the curves in (e-h).        
}
    \label{fig2}
\end{figure}

To quantify the patterns, we evaluate the spatial Fourier spectra for the surface height, $\zeta$, and surfactant concentration, $\Gamma$, defining $\hat{\zeta}_{mn}(t)$ and $\hat{\Gamma}_{mn}(t)$ to be the Fourier coefficients associated with the $(x,y)$ wavevector $\mathbf{k}_{mn}$. We then set $\zeta_{mn} \equiv \max_{\left[t,t+T\right]}{|\hat \zeta_{mn}(t)|}$ and $\Gamma_{mn} \equiv \max_{\left[t,t+T\right]}{|\hat \Gamma_{mn}(t)|}$.
Figures \ref{fig_2}(c,d) present an overview of the spatial Fourier spectra of $\zeta$ and $\Gamma$ as a function of $B$ in the range $B \in [0, 1.51]$. At higher $B$, ridges and hills emerge, where steep spatial gradients and many higher spatial harmonics appear. 

For $B < 1$, the square pattern is characterized by comparable amplitudes of $\zeta_{10}$ and $\zeta_{01}$, as shown in figure \ref{fig_2}(c). For $B < 0.5$, where Marangoni effects are weak, the $\zeta_{mn}$ modes have magnitudes similar to those associated with the clean case corresponding to $B = 0$, consistent with previous findings \citep{constante2021role}. For $B>1$, Marangoni-driven stresses dominate over inertial effects.  The square symmetry is broken, and by $B = 1.23$, $\zeta_{10}$ surpasses $\zeta_{01}$, with an increase in higher-order modes, such as the $\zeta_{20}$ mode. As $B$ increases further, strong $x-$dependent modes emerge, leading to a transition from asymmetric squares to stripes (see figure \ref{fig_2}(c)).

 A parallel change occurs in the $\Gamma$-spectrum. For $B < 1$, the surfactant is advected without being significantly hindered by Marangoni stresses, aligning the $\Gamma$-spectrum with the $\zeta$-spectrum, where $\Gamma_{10}$ and $\Gamma_{01}$ dominate (see figure \ref{fig_2}(d)). For $B>1$, $\Gamma_{10}$ begins to surpass $\Gamma_{01}$. Thus, $B \approx 1$ is a pivotal point in the dynamics, at which there is an equilibrium between the opposing mechanisms of advection-driven surfactant inhomogeneity and Marangoni-driven homogeneity.

We now turn to the formation of hills and ridges on the interface. Figures \ref{fig2}(a-d) illustrate the evolution of a small portion of the interface, color-coded by surfactant concentration. During the first half-cycle, the ridges rise, and the fluid and surfactant flow up from the troughs, advecting the surfactant to the apex of the ridge. Figures \ref{fig2}(e-h) show two-dimensional projections containing arc $s$, as indicated in figure \ref{fig2}(a). As the surfactant is advected towards the apex, a 
$\Gamma$-deficit (higher $\sigma$) is created at the trough.

The capillary force resulting from the $\Gamma$-deficit leads to the emergence of a bulb on the ridge, surrounded by a narrow region of negative curvature, which we call a \emph{neck}, and which is highlighted by a red spot on the interface in Figs. \ref{fig2}(e-h); $\Gamma$ accumulates at the ends of the ridge as shown in Figs. \ref{fig2}(a,b). Marangoni stresses along $s$ counteract the $\Gamma$-inhomogeneity caused by the surface advection. This is shown in figure \ref{fig2}(i), where $\tau > 0$  and $u_t < 0$ along the arc $s$. We call this a \emph{barrier}. This barrier rigidifies the surface during the first half-cycle, leading to $|\mathbf u_s| \approx 0$ at $t = 3T/8$, as shown in figure \ref{fig2}(b).
 
The negative vorticity along the surface in figure \ref{fig2}(f) indicates that $\tau$ opposes the surface advection. Due to this barrier,  a backflow develops on the surface from the apex towards the neck, as indicated by $u_t > 0$ in the inset of figure \ref{fig2}(j). This drives surfactants from the apex towards the neck.  Simultaneously, the accumulated surfactant at the ends of the ridge flows towards the neck due to a similar mechanism, as illustrated by the red arrows in figure \ref{fig2}(b). During this process, the midpoint of the ridge rises to form a bulb; see figure \ref{fig2}(c). By $t=T/2$, $\Gamma$ is maximal (so $\sigma$ is minimal) at the neck.

The accumulated surfactant causes Marangoni stresses, with distinct peaks of $\tau > 0$ and $\tau<0$ across the neck  (figure \ref{fig2}(k)). The barrier is now formed at the neck (shown as a white dotted region in figure \ref{fig2}(c)) where these stresses in the region between the apex and the neck begin to oppose the flow reversal at half-cycle. Meanwhile, surface tension decreases at the neck. 
As a result, the neck begins to reopen (see the streamlines in figure \ref{fig2}(g)) as is commonly observed in surfactant-laden neck reopening phenomena, discussed in detail in \cite{constante2021role}.

In the next half-cycle ($t \ge T/2$), the ridge begins to fall. However, the opposing Marangoni stress between the neck and the apex ($\tau < 0$ in figure \ref{fig2}(k)) slows the collapse of this region. This slower descent of $u_t$ ($du_t/ds \approx 0$) leads to the formation of the hill on the ridge. Meanwhile, at $t = 3T/4$, the region between the neck and the trough continues to fall more quickly than the hill. This accelerated fall is driven by the surfactant gradients towards the trough ($\tau > 0$ as shown in figure \ref{fig2}(l)) which, instead of opposing the bulk flow as before, now begin to support it \textcolor{black}{due to $u_t > 0$}. The presence of two high-vorticity regions (blue zones) along the interface in figure \ref{fig2}(h) is an effect of the two distinct roles of Marangoni stresses at the neck. As a consequence, new ridges develop while the hills of the previous cycle are still present, as seen in figure \ref{fig_2} at $t = 0$, $B=1.83$.
\begin{figure}
    \centering
    \includegraphics[width=0.55\linewidth]{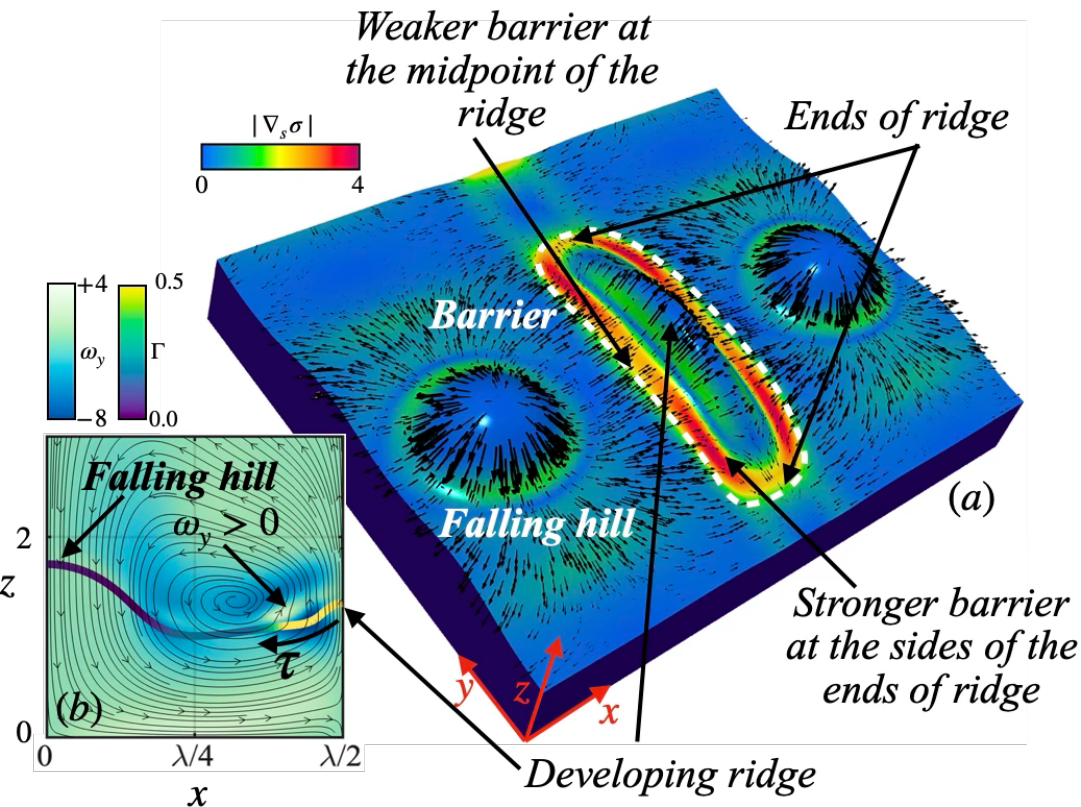}
     \captionsetup{width=1.0\textwidth}
    \caption{Marangoni-influenced ridge formation: (a) $x-z$ projection 
    containing
    $s$, as defined in figure \ref{fig2}, at $t = T$; the color-coding used here is that of figure \ref{fig2}. (b) Three-dimensional visualization of the interface color-coded by the magnitude of Marangoni stresses $|\nabla_s \sigma|$, indicating the barriers around the rising ridge.  
    }
    \label{fig3}
\end{figure}

Figure \ref{fig3} further elucidates the mechanism of ridge formation.
The surfactant accumulates on the developing ridge due to a combination of Marangoni-driven surface flow from the neck to the trough ($\tau>0$ in figure \ref{fig2}(h,l)) as previously discussed, and advection through bulk flow in the second half-cycle leading to strong surface compression at the ridge. This accumulation (see $\Gamma$-surplus region highligted in figure \ref{fig3}(a)) generates a Marangoni stress, directed from the newly developed ridge toward the falling hill (as highlighted by the arrow indicating the direction of $\tau$ in figure \ref{fig3}(a)). 
The magnitude of the Marangoni stress, $|\nabla_s \sigma|$, is shown in figure \ref{fig3}(b). This high-stress region, which surrounds the developing ridge, highlights the strength of the barrier to ridge formation. Close inspection of this region reveals that the barrier is weaker at the midpoint of the ridge, allowing stronger inward-directed surface flow to this region (viz. the velocity glyphs in figure \ref{fig3}(b)). This, in turn, leads to a higher elevation at the midpoint of the ridge than at its ends, as shown in figure \ref{fig3}(b). 

\section{Conclusion}
The study highlights the role of Marangoni stresses in Faraday wave pattern transitions. Numerical simulations were validated against previously reported two-dimensional simulation. 
Using the parameter $B$ to compare the Marangoni and inertial timescales, we found that the threshold acceleration increases with $B$. After we evaluated $B$, we increment the acceleration by 10\% of their respective threshold acceleration. Square patterns are observed for the surfactant-free interface. For the surfactant-covered interface, we found four different patterns as we increased $B$. We showed that at $B \approx 1$, square patterns transition to asymmetric squares. Increasing Marangoni strength further, asymmetric squares change to weakly wavy stripes. The novel finding highlighted here is the fact that at further higher $B$ values, ridges and hills appear. Due to strong Marangoni flow during a cycle of forcing, surfactant and surface flow compete ($\tau > 0$  and $u_t < 0$), which we call a barrier. The barrier slows down a rising ridge which then reaches its maximum height, resembling a bulb, at $t = 3T/8$. While the bulb falls in the next half cycle, a $\Gamma$-surplus region forms at the neck of the ridge. This creates a bi-directional Marangoni stress, where the flow is opposed (supported) between the apex (neck) and the neck (trough). This led to a faster collapse of the ridge between the neck and the trough. However, the bulb falls at a slower rate resembling a hill structure on the ridge. In the next cycle, $\Gamma$ accumulates at the newly forming crest. The barrier is weaker at the midpoint than at the sides of the rising crest. This creates a faster rise of the midpoint of the crest, resembling a ridge structure. The existence of such a barrier at the newly forming crest and at the neck are the cause of the formation of these interesting ridges and hills.


\backsection[Acknowledgement]{
This work was supported by the Engineering and Physical Sciences Research Council, UK, through the 
PREMIERE (EP/T000414/1) programme grant and the ANTENNA Prosperity Partnership (EP/V056891/1). O.K.M. acknowledges funding from PETRONAS and the Royal Academy of Engineering for a Research Chair in Multiphase Fluid Dynamics. D.P. and L.K. acknowledge HPC facilities provided by the Imperial College London Research Computing Service. D.J. and J.C. acknowledge support through HPC/AI computing time at the Institut du Developpement et des Ressources en Informatique Scientifique (IDRIS) of the Centre National de la Recherche Scientifique (CNRS), coordinated by GENCI (Grand Equipement National de Calcul Intensif) grant 2024 A0162B06721. The numerical simulations were performed with code BLUE \citep{shin2017solver,shin2018jcp} and the visualisations were generated using ParaView.}

\bibliographystyle{jfm}
\bibliography{jfm}

\end{document}